\documentclass[acmlarge,nonacm]{acmart}

\AtBeginDocument{%
  }

\copyrightyear{2026} 
\acmYear{2026} 
\setcopyright{rightsretained}

\acmConference[CHI '26 Workshop]{CHI Conference on Human Factors in Computing Systems Workshop: Body Transformation Experiences}{April 13--17, 2026}{Barcelona, Spain}
\acmBooktitle{CHI Conference on Human Factors in Computing Systems Workshop (Body Transformation Experiences: A workshop on How to Elicit, Assess and Support them through Multisensory Technology), April 13--17, 2026, Barcelona, Spain} 
\acmDOI{} 
\acmISBN{}

\makeatletter
\renewcommand{\section}{\@startsection{section}{1}{\z@}%
  {-3.5ex \@plus -1ex \@minus -.2ex}%
  {2.3ex \@plus.2ex}%
  {\reset@font\Large\bfseries}} 
\makeatother

\renewcommand{\refname}{\textbf{References}}

\begin{document}

\title{Reconciling Kinesthetic Mismatches: A Somatic Alignment Mindset for Musical Body Transformation}

\author{Ziyue Piao}
\email{ziyue.piao@mail.mcgill.ca}
\orcid{0009-0007-1281-9469}
\affiliation{%
  \institution{IDMIL, MPBL, CIRMMT, McGill University}
  \city{Montreal}
  \country{Canada}
}

\author{Isabelle Cossette}
\email{isabelle.cossette1@mcgill.ca}
\affiliation{%
 \institution{MPBL, CIRMMT, McGill University}
  \city{Montreal}
  \state{Quebec}
  \country{Canada}}

\author{Marcelo M. Wanderley}
\email{marcelo.wanderley@mcgill.ca}
\orcid{0000-0002-9169-4313}
\affiliation{%
  \institution{IDMIL, CIRMMT, McGill University}
  \city{Montreal}
  \state{Quebec}
  \country{Canada}}

\renewcommand{\shortauthors}{Piao et. al}

\begin{abstract}
Mastering musical performance requires precise multisensory coordination, yet learners encounter a kinesthetic mismatch, which is a discrepancy between the internal perception of an action and the actual physiological state of the body. While multisensory Body Transformation Experiences (BTE) provide tools to bridge this gap, existing designs often focus on external correction rather than internal alignment. To address this, we propose the Somatic Alignment Mindset (SAM), a conceptual lens that integrates Taoist philosophy to shift the focus of HCI design from prescriptive feedback toward holistic embodied unity. By positioning technology as a reflective medium, SAM operationalizes the principles of Adaptation, Assessment, and Awareness to reconcile somatic discrepancies and foster deep, self-aligned musical mastery.
\end{abstract}



\received{19 February 2026}

\maketitle
\begin{minipage}{\textwidth}
    \footnotesize
    \textbf{ACM Reference Format:}\\
    Ziyue Piao, Isabelle Cossette, and Marcelo M. Wanderley. 2026. Reconciling Kinesthetic Mismatches: A Somatic Alignment Mindset for Musical Body Transformation. In \textit{2026 CHI Conference on Human Factors in Computing Systems Workshop (Body Transformation Experiences: A workshop on How to Elicit, Assess and Support them through Multisensory Technology), April 13--17, 2026, Barcelona, Spain}. ACM, New York, NY, USA, 5 pages. 
\end{minipage}

\section{\textbf{Introduction}}
Mastering a musical instrument or singing is a complex multi-sensory task, requiring accurate physical coordination and mental awareness~\cite{zatorre2007brain}. Kinesthetic experiences, often described as the `feel' of bodily activity, are inherently internal and essential to learning motor skills in music performance~\cite{samudra2008memory}, yet they remain difficult for learners to monitor and control. For instance, singing learners must develop a kinesthetic awareness of breath control, the contraction of their respiratory muscles, and expansion of their chest wall~\cite{kleber2014neuroscience}, while piano students refine techniques by perceiving the kinesthetic feedback of finger and arm movements to achieve phrasing and dynamics~\cite{bangert2003mapping}. 

However, grounded in theories of motor control and sensory integration, we observe that learners often experience a `kinesthetic mismatch': a phenomenon we define as the discrepancy between the brain's internal representation of an action and the actual physiological state of the body~\cite{proske2012proprioceptive, schmidt2018motor}. For example, when a singing teacher instructs a student to increase physical energy for more even breathing, the student may visually mimic the teacher's outward gestures. Due to a kinesthetic mismatch, the student’s brain may misinterpret this increased energy as a requirement for isometric tension, leading to counterproductive muscle force and vocal strain rather than the intended breath support. Even after extensive training for hundreds of trials, adaptation tends to plateau because subtle kinesthetic errors may go undetected and may not be corrected~\cite{schmidt2018motor}. Prolonged or undetected mismatches may reinforce inefficient habits, hinder technical progress, and increase the risk of pain, injury, or frustration, sometimes causing students to quit their musical training prematurely~\cite{wolpert2000computational, wolpert1995internal, hansen2006common}.

Traditional teaching methods such as kinesthetic exercises, physical touch, and metaphor-based instructions rely on learners' ability to process and apply the kinesthetic information~\cite{piao2026want}. Although beneficial, these methods often fail to accurately capture complex bodily dynamics and yield inconsistent results across trials and individuals~\cite{ericsson1993role}. Recent advances in embodied and adaptive feedback technologies for Body Transformation Experiences (BTE), including wearable haptic devices, present an opportunity to make kinesthetic information both objective and actionable for music learners.

While multisensory signals, such as tactile sensations and haptic cues, are well-documented as foundational for sensorimotor learning in music, most existing BTE either lack the real-time dynamics necessary for self-calibration or fail to target the fundamental kinesthetic mismatch between internal perception and physical action~\cite{leborgne2024vocal, hirano2025enhanced}. Haptic devices offer a transformative pathway to augment the perception of complex physical movements~\cite{tom2020haptic} by enabling students to externalize and reflect upon their own internal physiological states. Previous research demonstrates that wearable haptics can a) elicit deeper kinesthetic awareness of breath control in singers~\cite{lu2025project}, b) provide tactile anchors for refining rhythmic and motor skills~\cite{lederman2003neuromuscular}, and c) facilitate a shared somatic dialogue between teacher and student~\cite{piao2023mappemg}. However, these interventions rarely resolve deep-seated kinesthetic mismatches, often remaining confined to laboratory settings with non-portable equipment. 

We propose the Somatic Alignment Mindset (SAM) as a potential design lens for BTEs in music learning contexts. By outlining three core strategies (Adaptation, Assessment, and Awareness), we aim to provide a conceptual foundation for addressing kinesthetic mismatches in future somatic interactions. This approach integrates Taoist philosophy to shift the focus of Human-Computer Interaction (HCI) design from corrective feedback toward holistic embodied awareness. SAM positions technology as a reflective medium that augments the user’s natural learning process. Furthermore, we aim to engage the BTE design community to identify real-world mismatches, using these practitioner insights to iteratively refine and ground the SAM lens in practical application.

\section{\textbf{Somatic Alignment Mindset}}
Somatic Alignment Mindset (SAM) that introduces Taoist philosophy to reconcile kinesthetic mismatches. This mindset suggests treating BTEs into a reflective medium, guiding learners from sensory-motor discrepancy toward a state of embodied unity, an experiential ``Embracing the One'' (Bao Yi) where internal intention and physical execution merge~\cite{zi2017dao}. BTEs are viewed not as external tools, but as somatic resonance partners that support the cultivation of a seamless, unified flow between the practitioner and their environment. We operationalize SAM through three core strategies: Adaptation, Assessment, and Awareness.

\begin{figure*}[h!]
  \centering
  \includegraphics[width=0.85\linewidth]{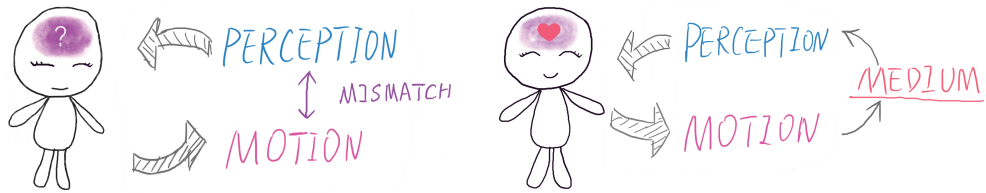}
  \caption{From Kinesthetic Mismatches to Bao Yi}
  \label{fig:mismatch}
\end{figure*}

\subsection{\textbf{Adaptation: \textit{Zi Ran} and the Logic of \textit{Wu Wei}}}
The first pillar of our roadmap focuses on the interaction logic between the body and the system. The strategy of Adaptation operationalizes the Taoist principles of \textit{Zi Ran} (Naturalness) and \textit{Wu Wei} (Effortless Action)~\cite{zi2017dao}  as distinct design constraints. In this context, \textit{Zi Ran} dictates that the sensory modalities of technology must conform to the body's innate kinetic logic, respecting physiological boundaries to ensure the interface feels like an organic extension rather than an alien attachment.

Parallelly, \textit{Wu Wei} (Effortless Action) informs the temporal logic of intervention. Rather than treating the user as a passive receiver of corrective commands, \textit{Wu Wei} prioritizes implicit, non-striving alignment that preserves the user's flow and agency. Instead of discrete ``correct'' or ``incorrect'' notifications, the system maintains a liminal presence—analogous to a reactive shadow that only manifests when the performer drifts from their somatic center. For instance, a piano learner does not receive a sharp alert for excessive tension; rather, they feel a subtle, growing texture via vibration that mirrors their forearm muscle effort~\cite{piao2023mappemg}. This ensures that the intervention is not perceived as an intrusive ``external shock,'' but as a self-originating somatic feedback of the misalignment itself. By transitioning from rigid correction to resonant guidance, BTEs can act as a silent partner, enabling the learner to achieve mastery through intuitive internal recalibration rather than cognitive rule-following.

\subsection{\textbf{Assessment: \textit{Nei Guan} as Reflective Somatic Alignment}}

Building upon adaptive interaction, Assessment is redefined as a process of navigating one's own internal landscape rather than a top-down ``diagnostic'' measurement. This shift is informed by the Taoist practice of Nei Guan (Internal Observation), which replaces the external ``judge'' with a state of Guan, which is a first-person non-judgmental, contemplative awareness of one's physiological state~\cite{zi2017dao}. Many BTE systems' evaluation prioritized quantitative accuracy, but we argue for a shift toward reflective feedback, where BTEs serve as interactive probes for supporting self-observation.

By treating wearables as bodily reflective tools, we transform physiological data into a ``somatic mirror'' that augments a pianist's perception of their internal body. Similar to how a stethoscope amplifies internal sounds to aid intuition, our Breathing Mirror systems provide a window into the learner's pre-reflective sensations~\cite{piao2026breathing}. This allows the participant to ``see'' their bodily dynamics or postural shifts as expressive data points and assess and reflect by themselves. Through this journey of internal observation, the SAM lens enables learners to reconcile kinesthetic mismatches by fostering a recursive somatic dialogue by turning physiological data into a meaningful conversation with their own body.

\subsection{\textbf{Awareness: Cultivating \textit{Bao Yi} as the Transformative Goal}}

The ultimate goal is the cultivation of Awareness, grounded in the Taoist concept of Bao Yi (Embracing the One). This represents the final state of mind-body unity where the kinesthetic mismatch is fully reconciled. While contemporary Western philosophy, from Merleau-Ponty's phenomenology \cite{merleau2013phenomenology} to Shusterman's somaesthetics \cite{shusterman2008body}, has long moved beyond Cartesian dualism, the core challenge lies in translating the philosophy of `oneness' into actual interface mechanics. Current BTE designs may reinforce a functional split by treating physiological data as an external object of analysis. By introducing the Taoist concept of Xin (Heart-Mind), we do not seek to re-establish a unified theory, but to provide a design vocabulary that treats the learner's emotional intent and kinetic execution as a singular, resonant system.

From a BTE design perspective, this culminates in sensory transparency, where the system functions as a sensory amplifier that translates physiological data into intuitive, cross-modal signals. By mapping subtle internal dynamics onto ambient modality textures such as haptic or auditory, the interface becomes a ``somatic resonator.'' This transparency allows the learner to perceive their internal state as a unified whole. At this stage, the technology is no longer an external monitor or even a tool, but an integrated part of the user's own kinesthetic loop. This represents the peak of the BTE: a journey where technical training dissolves into a pursuit of unified embodied awareness, and the practitioner and the technology exist in a state of harmonious ``Oneness'', which humans may hardly achieve by themselves.

\section{\textbf{Discussion}}

To provoke a deeper dialogue within the BTE community, we propose the following questions that challenge current design paradigms through the lens of SAM:

\begin{enumerate}
\item \textbf{Identifying and Addressing Kinesthetic Mismatch:}
The core of SAM is the reconciliation of Kinesthetic Mismatch. We invite participants to interrogate their own systems:
\begin{itemize}
\item \textbf{Locating the Gap:} Within your specific BTE intervention, where do the kinesthetic mismatches between the user's felt-sense and the system's feedback manifest? Is it a temporal lag, a spatial discrepancy, or a deeper semiotic conflict between the sensory signal and the user's intent? How do you capture these mismatches?
\item \textbf{Reductive vs. Productive Mismatch:} Does your system attempt to solve these discrepancies to provide a seamless experience, or could we leverage such mismatches as a creative friction? Can we design BTEs that allow practitioners to explore the productive mismatches in their proprioception rather than correcting them?
\end{itemize}

\item \textbf{Critical Reflections on the SAM Lens:} 
By applying Taoist principles to multisensory design, we open new questions regarding agency and the trajectory of transformation:
\begin{itemize}
    \item \textbf{The Paradox of Effortless Action:} If a BTE system becomes perfectly adaptive and effortless, at what point does \textit{Adaptation} turn into a ``somatic crutch'' that prevents the user from internalizing the transformation and developing their own \textit{Xin} (Heart-Mind)?
    \item \textbf{The Exit Strategy for Unity:} If the ultimate goal of a BTE is the cultivation of unified \textit{Awareness}, what is the technology's ``exit strategy''? Should the system be designed to eventually become obsolete once alignment is internalized, or is the future of BTE a permanent technological organ for the self?
\end{itemize}
\end{enumerate}

\begin{acks}
This research was supported by McGill University's Input Devices and Music Interaction Laboratory (IDMIL) and FRQSC Doctoral Scholarship (No. 0000342567) from Fonds de recherche du Québec. Thanks to Dr. Theodora Nestorova for her invaluable collaboration and for providing the creative inspiration behind the SAM framework. The authors express their sincere gratitude to Yohei Wada, Dr. Akira Maezawa, Dr. Katsunori Suzuki, Naoto Kojima, and other colleagues at Yamaha for their essential technical support and collaboration in wearable design, which is an important application of the framework. Special thanks also to the members of IDMIL and McGill's Music Performance and Body Lab for their invaluable feedback and support throughout this research.
\end{acks}

\bibliographystyle{ACM-Reference-Format}
\bibliography{sample-base}

\end{document}